\documentclass
[aps,prd,twocolumn,superscriptaddress,altaffilletter,lengthcheck,tightenlines,showpacs,showkeys]{revtex4-1}
\usepackage[dvipdf]{epsfig}
\usepackage{color}
\usepackage{sidecap}
\usepackage{graphicx}
\usepackage{subfigure}
\usepackage{amsmath}
\usepackage{amsfonts}
\usepackage{amssymb}
\newcommand{\be}{\begin{equation}}
\newcommand{\ee}{\end{equation}}
\newcommand{\n}{\label}
\newcommand{\no}{\noindent}
\newcommand{\ben}{\begin{eqnarray}}
\newcommand{\een}{\end{eqnarray}}
\usepackage{hyperref}
\begin{document}

\title{Linking phantom quintessences and tachyons}

\author{M\'onica Forte}
\affiliation{Departamento de F\'isica, Facultad de Ciencias Exactas y Naturales,
Universidad de Buenos Aires, 1428 Buenos Aires, Argentina,\\forte.monica@gmail.com}

\begin{abstract}
We present a nonparametric form-invariance transformation through which we establish a link between phantom and tachyonic models in flat Friedmann-Robertson-Walker cosmologies.

\end{abstract}
\pacs{98.80.-k,03.50.-z,04.25.-g,11.30.-j}

\keywords{tachyon, phantom quintessence, symmetry}

\maketitle

\section{Introduction}

       The concept of symmetry has an undoubted importance in physics, both in the dynamical cases and in nondynamical ones \cite{Glashow:1961tr,Cabibbo:1963yz,'tHooft:1976up,Higgs:1964ia,Susskind:1978ms,Arnold:1989,G. Chavchanidze,LMRañada,Marsden-Weinstein,Ranada1,Ranada2,Sarlet-Cantrijn,Gordon:1984xb}. Dynamical symmetries correspond to some inherent property of the matter or spacetime evolution and constrain the solutions of the equations of motion while  nondynamical symmetries arising because of the way in which we formulate the action and give rise to mathematical identities. Examples of the latter are the non-Noether symmetries \cite{Noether:1918zz,Lutzky,FuChenJimenezTang,FuChen,Lunev,Djukic:1980ya} and the form invariance symmetries \cite{Chimento:1997uj,YiMeiMei,Mei,Cai:2010zz,Chimento:2002gb,Chimento:2003qy,Chimento:2003ie,Cataldo:2005gb,ChimentoAcad,Chimento:2007fx,Chimento:2008pj,Chimento:2009rw}.

The transformations that respect the form invariance symmetries (FIT) are very useful in several interesting cosmological problems not only to linearize complicated equations of motion but also to generate new cosmologies from the old known inside the same metric \cite{Chimento:2012qx}. In this paper we will devote attention to a nonparametric group of  FIT that leaves invariant the Einstein equations like identities. 

The Einstein equations for a flat Friedmann-Robertson-Walker (FRW) cosmological model with factor of scale $a$, filled with a perfect fluid with energy density $ \rho$ and pressure p, are
\begin{subequations}
\n{1}
\ben
\n{einsteina}
3 H^2 = \rho, \qquad H=\dot a/a,
\een
\be
\n{conserv}
\dot \rho+3H(\rho+p)=0,
\ee
\end{subequations}
where a dot denotes differentiation with respect to the cosmic time.

Another  cosmological model filled with a perfect fluid which energy density is $\bar\rho$ and pressure is $\bar p$, will generate similar expressions
\begin{subequations}
\n{2}
\ben
\n{bar einsteina}
3 \bar H^2 =\bar\rho,  \qquad \bar H=\dot{\bar{a}}/\bar{a},
\een
\be
\n{bar einsteinb}
\dot{\bar\rho}+3\bar H(\bar\rho+\bar p)=0.
\ee
\end{subequations}

Form invariance concept such as used in \cite{Chimento:2002gb}, refers to  nonlocal transformations which connect the two sets (\ref{1}) and (\ref{2}) leaving unchanged those expressions. This implies that the transformations, described as connections between the energy densities, must meet the requirements
\begin{subequations}
\n{3}
\be
\n{rhoa}
\bar\rho=\bar\rho(\rho),
\ee
\be
\n{rhob}
\bar H= \Big(\frac{\bar\rho}{\rho}\Big)^{1/2} H,
\ee
\be
\n{rhoc}
\bar p= -\bar\rho +\Big(\frac{\rho}{\bar\rho}\Big)^{1/2}(\rho+p) \frac{d\bar\rho}{d\rho}
\ee
\end{subequations}
\no for any invertible function $\bar\rho(\rho)$.

If we consider perfect fluids with barotropic equations of state $p=(\gamma - 1)\rho$ and $\bar p=(\bar\gamma - 1)\bar\rho$, respectively, it follows that the barotropic indices are related by
\be
\n{indices}
\bar \gamma= \Big(\frac{\rho}{\bar\rho}\Big)^{3/2}\frac{d\bar\rho}{d\rho}  \gamma.
\ee


\section{The nonparametric group $\mathcal{G}$}

Consider the nonparametric  group of FIT $\mathcal{G}=\{\mathbb{I},\mathbb{F}\}$ composed by the identity $\mathbb{I}(\rho)=\rho$ and the transformation  $\mathbb{F}(\rho)=\rho^{-1}$. The former trivially satisfies the prescriptions (\ref{3}) while the latter results in the expressions
\begin{subequations}
\n{5}
\be
\n{transf1}
\bar H= \frac{1}{3H},
\ee
\be
\n{transf2}
\bar p= -\frac{1+\rho+p}{\rho},
\ee
\be
\n{barotr}
\bar\gamma=-\gamma\rho
\ee
\end{subequations}

\no for the transformed Hubble factor, global pressure, and barotropic index. Moreover, from (\ref{transf1}) and $H = d \ln a / dt$ the transformed factor of scale is expressed as
\be
\n{factor}
\bar a= \bar a_0 \exp\Big(\int \frac{da}{a\rho(a)}\Big).
\ee

Equation (\ref{factor}) allows us to draw some features about models connected by this FIT. For example, looking at the power law solutions of (\ref{1}) $\rho=\rho_0 a^{-n}$, it is concluded that de Sitter universes are connected with de Sitter universes if $n = 0$. When  $n \ne 0$, the connection is more interesting. 
For $n < 0$, contractive super accelerated  models ($\gamma < 0$) are connected with contractive models that exhibit  decelerated-accelerated transition at some time $ t_{trans} =-6/n > 0 $ but they are never super accelerated (because $\bar\gamma =-4/nt^2 > 0$ for them).  If $n > 0$, expansive models always accelerated ($0 < n < 2$) or always decelerated ($2 < n $) are linked with expansive super accelerated models.
Note that for power law solutions, this FIT never connects a universe in expansion with a contracting one. Therefore, these dualities are different in nature to the old known $a \  \leftrightarrow \  a^{-1}$ duality \cite{Meissner:1991zj,Veneziano:1991ek,Sen:1991zi,Lidsey:1995ft,Chimento:2003qy,Cai:2006dm}.

We can still note another interesting characteristic from (\ref{factor}) when the $\Lambda$CDM model $\rho=\Lambda +\rho_1 a^{-3} $ is considered as the object of the transformation. The bouncing solution 
\begin{equation*}
\n{lcdm}
a= \left(\sqrt{\frac{\rho_1}{\Lambda}}\sinh\left[\sinh ^{-1}\Big(\sqrt{\frac{\Lambda a_0^3}{\rho_1}}\Big)+\frac{\sqrt{3\Lambda}}{2}(t-t_0)\right]\right)^{2/3}, 
\end{equation*}
\no which is singular for $t_b=t_0-\frac{2}{\sqrt{3\Lambda}}\sinh ^{-1}(\sqrt{\Lambda a_0^3/\rho_1})$, is mapped onto the transformed, nonsingular bouncing solution

\begin{equation*}
\n{barlcdm}
\frac{\bar a}{\bar a_0}=\frac{\left(\cosh\left[\sinh ^{-1}\Big(\sqrt{\frac{\Lambda a_0^3}{\rho_1}}\Big)+\frac{\sqrt{3\Lambda}}{2}(t-t_0)\right]\right)^{2/3\Lambda}}{(1+\frac{\Lambda a_0^3}{\rho_1})^{1/3\Lambda}},
\end{equation*}
showing a new type of duality, between singular and nonsingular bouncing models.


\subsection{Linking phantom and tachyonic models}

Finally, consider a FRW model,  filled with a phantom quintessence field $\varphi$  with energy density $\rho_{\varphi}$, pressure $p_{\varphi}$, and barotropic index $\gamma_{\varphi}$,
\begin{subequations}
\n{varphi}
\be
\n{rhovarphi}
\rho_{\varphi}=-\frac{\dot\varphi^2}{2} + U(\varphi),
\ee
\be
\n{pvarphi}
p_{\varphi}= -\frac{\dot\varphi^2}{2} - U(\varphi),
\ee
\be
\n{gammavarphi}
\gamma_{\varphi}= \frac{-2\dot\varphi^2}{-\dot\varphi^2 + 2 U(\varphi)},
\ee
\end{subequations}
\no with  $U(\varphi)$ an arbitrary positive phantom potential. The transformation $\mathbb{F}(\rho)$ allows us to consider the  $\varphi$ field  as a tachyon $T$ field  with energy density $\rho_T$, pressure $p_T$, and barotropic index $\gamma_T$,
\begin{subequations}
\n{tach}
\be
\n{rhotach}
\rho_T=\frac{V(T)}{\sqrt{1 - \dot T^2}},
\ee
\be
\n{ptach}
p_T= -V(T)\sqrt{1 - \dot T^2},
\ee
\be
\n{gammatach}
\gamma_T=  \dot T^2
\ee
\end{subequations}
because (\ref{barotr}) equalizes both fields (to less than a constant) if the tachyonic model is considered as the transformation of the phantom quintessence one. Therefore, this transformation describes a duality between the tachyonic and phantom descriptions  of a cosmological model in a flat FRW metric. The phantom Lagrangian $\mathcal{L}=-\dot\varphi^2/2 - U(\varphi)$ is dual, under the FIT  $\mathbb{F}(\rho)$, to the  tachyonic Lagrangian $\mathcal{\bar L}= -V(T)\sqrt{1 - \dot T^2}$, where the tachyonic potential is obtained by the expression   
\be
\n{UentV}
V(T)= \frac{\sqrt{1 - \dot\varphi^2(T)}}{U(T)- \dot\varphi^2(T)/2},
\ee
\no and the constant difference between $\varphi$ and $T$ is taken as zero. Conversely, given the tachyon potential $V(T)$, the phantom potential $U(\varphi)$ is obtained by the expression
\be
\n{VentU}
U(\varphi)= \frac{\dot T^2(\varphi)}{2}+ \frac{1}{V(\varphi)}\sqrt{1 - \dot T^2(\varphi)}.
\ee

As a brief example, consider a FRW model driven by a phantom quintessence scalar field $\varphi$, with the self-interaction $U(\varphi)={\dot \varphi}^2/2+\lambda \varphi^2$ \cite{Chimento:2008ws}. This form of potential supposes that $\dot \varphi(\varphi)$, and in this case it is related with power law potentials that have been extensively considered for the inflaton field and for chaotic models \cite{Turner:1983he,Tsujikawa:2000tm,delaMacorra:2002bp,Martin:2010kz,Tong:2013rs}

From Eq.  (\ref{rhovarphi}), it turns out that this potential leads to the relationship $H=\sqrt{\lambda/3}\ \varphi$ and from the continuity equation for the phantom, it follows that $\varphi(t)=\sqrt{4\lambda/3}\ t +\varphi_0$. Therefore, $U(\varphi)=\lambda \varphi^2 +2\lambda/3$ is a quadratic potential and
the factor of scale has the expression $a(t)=a_0exp(\lambda t^2/3 +\sqrt{\lambda/3}\varphi_0\ t )$. 
This phantom model is linked with a  FRW universe commanded by a tachyon field $T(t)\equiv \varphi(t)$ subjected to an inverse square tachyonic potential $V(T)= \sqrt{1 - 4\lambda/3}/\lambda T^2$ with a factor of scale $\bar a/\bar a_0= (\sqrt{4\lambda/3}t +\varphi_0)^{1/2\lambda}$.

Note that this connection is very different from those obtained in 
\cite{Avelino:2010qn} and \cite{Avelino:2011ey}, where both descriptions, the scalar and the tachyonic, have the same background evolution but differ at the perturbative level (see also \cite{Piao:2004uq} for dualities of primordial perturbation spectrums).
 Here the evolution is different at every level, even though the fields that command each scenario always have the same temporal evolution.

\section{Conclusions}

Summarizing, we have shown that for power law solutions of (\ref{1}) with energy densities of the form $\ \rho=\rho_0 a^{-n}$, there are links between (i) always super accelerated, contractive universes with contractive universes that suffer a transition between nonaccelerated and accelerated regimes for $n < 0$, (ii) de Sitter models with de Sitter models for $n = 0$, and (iii) always accelerated or always decelerated expansive models with super accelerated expansive models for $n > 0$. Also, the singular bouncing solution of the $\Lambda$CDM model is linking with a nonsingular bouncing solution.  Most important, for flat FRW cosmologies, the tachyonic universes are dual to phantom quintessence  models. As a matter of fact, these dualities are particular cases of the form-invariance symmetry of the corresponding gravitational field equations under the nonparametric group $\mathcal{G}$ .

\section{ACKNOWLEDGMENTS}
We would like to thank the anonymous referee for making useful suggestions which helped improve the paper.


\end{document}